\documentclass[a4paper]{article}

\usepackage{16lomcon}        
\usepackage{cite}             
\usepackage{epsfig}           

\newcommand{\be}{\begin{eqnarray}}
\newcommand{\ee}{\end{eqnarray}}
\newcommand{\bi}{\bibitem}

\newcommand{\rar}{\rightarrow}

\begin{document}


\title{LIGHT MILLICHARGED PARTICLES AND LARGE SCALE COSMIC MAGNETIC FIELDS}

\author{Alexander D. Dolgov \email{dolgov@fe.infn.it}
}

\affiliation{
 Novosibirsk State University, Novosibirsk, 630090, Russia\\
 Institute of Theoretical and Experimental Physics, Moscow, 113259, Russia \\
Dipartimento di Fisica, Universit\`a degli Studi di Ferrara, I-44100 Ferrara, Italy \\
}


\date{}
\maketitle


\begin{abstract}
After a brief review of different types of scenarios suggested for the solution of the problem of galactic and
intergalactic magnetic field generation, the mechanism based on the electric  current induction by hypothetical 
millicharged particles interacting with electrons in cosmic medium is discussed. The proposed model successfully
describes observational data. The new light millicharged particles can contribute from a small fraction up to
100\% to the cosmological dark matter.

 \end{abstract}

There are strong observational observational evidence for  existence of galactic magnetic fields
with the field strength  ${ B_{gal}}$  of a few ${\mu G}$ and the coherence length of
 several kiloparsecs. There are also some data in favor of weaker intergalactic magnetic fields, {${ B \sim 10^{-9} }$ G,} 
 but with much larger coherence scale of hundreds kiloparseks.  For a review on observations see refs.~\cite{B-obs-rev}.
 The origin of such large scale  galactic and  intergalactic magnetic fields remains a cosmological mystery for many 
 years.

There are a lot of proposals for mechanisms of large scale magnetic field generation. They can be roughly divided into 
two classes: either ones based on conventional physics, or those which invoke different new physics hypotheses.  
They are reviewed e.g.  in papers~\cite{theor-rev}. Both types of mechanisms suffer from the similar problems. The 
magnitude of the generated field might be quite high but the coherence length happened to be much shorter than
the necessary one, or vice versa: the coherence length is large but the magnitude is tiny. The problems are milder in
the case of new physics allowed but they still remain vital. The case of tiny but the galactic scale seed magnetic field to 
some extend could be  cured by the galactic dynamo amplification. However, the seed fields are usually so weak that 
the galactic dynamo is not sufficient  to amplify them up to the observed magnitude. 
As for intergalactic fields, the dynamo practically is not efficient there, so the existence of intergalactic fields puts strong  
restrictions on the mechanisms of generation of primordial (seed) magnetic fields. 

It is tempting to search  for a possibility to generate primordial magnetic fields during inflation, since the problem of
large scale in this case is naturally and easily solved by exponential stretching. However,  an electromagnetic field cannot be
created during inflation because the classical electrodynamics is conformally invariant and such fields are not produced 
in conformally flat Friedmann-Robertson-Walker metric~\cite{parker}. However, quantum triangle anomaly breaks conformal
invariance and allows for photon production and thus for magnetic field generation~\cite{ad-conf-anom}. 
This is a promising mechanism but more detailed investigation of renormalization in De Sitter space-time is necessary.
Another, more exotic possibility, is an assumption of non-minimal coupling of electromagnetic field  to gravity through a
new interaction, e.g. by  term ${\xi R A_\mu A^\mu}$~\cite{turner-widrow}, where $R$ is the curvature scalar and $\xi$ is a constant.
Such an interaction which not only breaks conformal but the gauge  invariance of electrodynamics as well. As a result
the photon would acquire non-zero mass, due to one-loop graviton-photon diagram,
of the order $m_\gamma \sim \xi \Lambda^2 /m_{Pl} $, where $\Lambda$ is an
ultraviolet cut-off and $m_{Pl} \approx 10^{19}$~GeV is the Planck mass. An existence of the galactic magnetic fields 
implies $m_\gamma < 1/$kpc, which in turn demands $\Lambda < 10\,$eV$/\xi$. So for a reasonable ultraviolet cutoff 
a tiny $\xi$ is necessary which would drastically diminish the effect of this new interaction on generation of primordial
magnetic field.

At later cosmological stages very strong magnetic fields could be created at first order phase transitions but the coherence
length in this case would be microscopically small and even with cosmological expansion it would remain by far too short. 
This restriction is avoided if e.g. at electroweak phase transition a condensate of electrically charged W-bosons were
formed~\cite{ad-al-gp}.  In this case the primeval plasma could be spontaneously magnetized inside macroscopically 
large domains  and such magnetic fields may be the seeds for the observed today galactic and intergalactic fields.

Next interesting period for generation of primordial magnetic fields might be the epoch of big bang nucleosynthesis
at ${t \geq 1}$ sec, ${T \leq }$ 1 MeV. The coherence scale in this case is bounded form above  ${ l_c < 100}$  pc in terms 
of the present day units. In the case of big and inhomogeneous primordial lepton asymmetry the magnitude of the
the created magnetic field may be sufficiently large, such that after chaotic field line reconnection (analogous to the Brownian 
motion) the coherence scale might extend to galactic, but not to intergalactic scales~\cite{ad-grasso}. This process proceeds 
at the expense of a decrease of the field amplitude and galactic dynamo is very much in order.

During the last decade an attention was attracted to generation of cosmic magnetic field
during or around the epoch of recombination, ${z \sim 1000}$, or even closer to the present day.
at the period of the large scale structure formation. Magnetic field can be produced by vortex 
currents, but vorticity perturbations are absent in the primordial density perturbations. However, as
it was found in refs.~\cite{zb-ad}, vorticity may be created by the photon diffusion in the second 
order in the usual scalar, non-vortex, perturbations. 

Still in this case and other cases mentioned above a huge, sometimes even unrealistic, dynamo amplification 
by galactic rotation is  necessary.  This could help to amplify galactic magnetic fields from originally weak seed 
fields, but fails for intergalactic magnetic field.

In our recent work~\cite{zb-ad-it} we suggested a new mechanism of magnetic field generation in the universe with
already formed first large scale structures, assuming that there exist light millicharged particles.
Let us consider first a protogalaxy rotating in the background of the CMB photons.
The force exerted by photons on electrons is much larger than that exerted on protons,
${F\sim \sigma \sim 1/m^2}$, where $\sigma$ is the cross-section of elastic $\gamma e $ or $\gamma p$ scattering.
The difference between $e$ and $p$  accelerations is even bigger, because ${a \sim F/m \sim 1/m^3}$.
{So a circular electric current, proportional to the rotational velocity of the protogalaxy, ${v_{rot}}$, must be induced.} 
The force acting on electrons is given by the expression:
$\vec{F} \sim  \vec{v} \sigma_{e\gamma}^{~} n_\gamma \omega_\gamma$. This force coherently acts on electrons
during  the collision time determined by $ep$-scattering, which is equal to:
\begin{equation} 
\tau_{ep} = \frac{m_e^2 \langle v_e^2 \rangle }{4\pi  \alpha^2 \langle 1/v_e \rangle n_e L_e} 
\simeq  \frac{m_e^{1/2} T_e^{3/2}}{ 4\pi \alpha^2 n_e L_{e}} \, ,
\end{equation}
where the Coulomb logarithm is ${L_e  \sim 10}$  and ${\langle v_e^2 \rangle = T_e/m_e}$. So 
we obtain the standard expression for the conductivity: 
$ \kappa = {e^2 n_e  \tau_{ep}}/{2 m_e}   
\simeq    {T_e^{3/2} }/{  8\pi \alpha  L_e m_e^{1/2}} . $
The conductivity
does not depend on the density of the charge carriers, ${n_e}$, 
{unless the latter is 
so small that the resistance is dominated by neutral particles.  }

{Thus the difference between rotational velocities of ${e}$ and ${p}$ is}
{${\Delta v_e = \tau_{ep} F /2m_e }$}
{and the current} {${j = e n_e \Delta v_e}$. } Naively estimating ${B}$ by the
Biot-Savart law as {${B \sim 4\pi j R } $} 
where ${R}$ is the galaxy radius, we find that
for a typical galaxy with ${R\sim 10}$ kpc 
{${v_{\rm rot} \sim 100}$ km/s:} 
$ B \sim \mu  $G,  {very close to the observed value without any dynamo.}
However, this is incorrect since the time to reach stationary (Bio-Savart) limit is longer than
the cosmological time.

To proceed further we need to use the MHD equation modified by presence of an external force:
\be 
 \partial_t \vec B = 
 {{\vec{\nabla} \! \times \! \vec{F}/e \, + \, \vec{\nabla} \! \times \! (\vec{v} \! \times \! \vec{B}) \, + }} 
 { {  (4\pi\kappa)^{-1} (\Delta \vec B -  \partial^2 _t \vec B). } }
\ee
In the limit of high conductivity, the second term in the
equation, the advection term, can lead to 
dynamo amplification of magnetic seed fields once the value
of the latter is non-zero. 

Assuming ${B=0}$ at ${t=0}$, we find 
$\vec B(t) =  \int_{0}^t  dt \, \vec \nabla \! \times \! \vec F/e  $.
The largest value of the magnetic seed is generated 
around the hydrogen recombination at ${z_{\rm rec} \sim 10^3}$, 
or {${t_{\rm rec}  \sim 5 \times 10^5}$~yr. } 
{Earlier  the plasma was strongly coupled and the relative motion of electrons and protons was negligible. }

The seed field generated at this epoch with  coherence length  {${\lambda \sim 1}$ kpc,} 
corresponding to the present scale of a typical galaxy ${\sim 1}$ Mpc, is very weak
\be 
B_\lambda \sim  \Omega_\lambda t_{\rm rec} B_F(t_{\rm rec}) \leq 10^{-20}\,G,
\ee
where {$ {\Omega_\lambda = | \vec\nabla \! \times \! \vec v |_\lambda  \leq 10^3(\delta T/T)^2/\lambda}$.}
{The seeds with the coherence length of a few kpc and  
{${B_{\rm seed} >  10^{-15}}$~G} are needed to fit the observations. }

Now we consider the case when instead of CMB photons new DM particles, ${X}$ collide with protons and electrons
and similarly generate an electric current. The current is 
proportional to the cross-section of ${Xe}$-elastic scattering, ${\sigma_{Xe}}$,  to ${n_X/n_e}$ ratio,
 and to   X-particle momentum, {${p_X  = m_X v_{rot} } $.} 
Therefore, to produce stronger than CMB force on electrons,  {${\sigma_{Xe}}$
should be large.} {This is possible if ${X}$ have long range interaction leading to an enhancement of  
${\sigma_{Xe}}$ at low momentum transfer. } So we consider millicharged particles with the mass from 
a few keV  to several MeV.

The bounds on the  X-particle charge, ${e' = \epsilon e} $ are summarized in ref.~\cite{zb-ad-it}.
{If ${m_X < m_e}$, then
from ortho-positronium invisible decays follows ${\epsilon < 3.4\cdot 10^{-5} }$.}
{For ${m_X = 1}$ MeV:  ${\epsilon < 4.1 \times 10^{-4}}$.} 
{For ${m_X=100}$ MeV: ${\epsilon < 5.8 \times 10^{-4}}$.}
{We assume that  ${m_X > 10}$ keV to avoid strong limits on ${e'}$  from the
 stellar evolution. }
{The BBN bounds can be relaxed if the lepton asymmetry is non-zero~\cite{zb-ad-it-bbn}. }

If X-particles were thermally produced, their abundance could be calculated according to the Zeldovich~\cite{zeld-eq} 
(a decade later called Lee-Weinberg) equation:
\be 
\Omega_X h^2   \approx 0.023 ~ x_f \, g_{\ast f}^{-1/2} 
\left(\frac {v\sigma_{\rm ann}}{ 1~{\rm pb} }\right )^{-1}, 
\ee
where $x_f \equiv {m_X}/{T_f} = 10 +{\rm  ln} [(g_X \, x_f^{1/2} m_X) /9{g_{\ast f} \, MeV )]}$,
${g_X}$ is the number of the spin states  of X-particle, and  ${ g_{\ast f}}$ is the 
effective number of particle species in the plasma at ${T=T_f}$.
{If ${m_X < m_e}$, X-particles can annihilate only into  photons with }
 $ v \sigma (X \bar X \rar 2\gamma) = {\pi \alpha^{\prime 2} }/{ m_X^2}$,
where $ { \alpha^\prime = e^{\prime \, 2} /4\pi = \epsilon^2 \alpha } $. 
Thus
$\Omega_X h^2   \approx 150\, \left(  { 10^{-5} m}/{\epsilon^2\, \rm{keV} } \right)^2$.
Hence X's would be overproduced if
${\epsilon < 3.4\cdot 10^{-5} }$. Additional  
annihilation into ${\bar \nu \nu}$ or dark photons could help, even if the CMB constraint 
is fulfilled: ${\Omega_X h^2 < 0.007}$~\cite{dgr}.

If {$m_X >m_e$, then the channel ${X\bar X \rar e^+e^-}$ is open and
$v \sigma (X \bar X \rar e^+e^-) = {\pi \alpha \alpha' }/{ m_X^2}$.
Correspondingly:
$\Omega_X  h^2 = 0.012 \, \left( { 10^{-5} m}/{\epsilon\, {\rm MeV}} \right)^2 $ 
and e.g. for  ${m_X = 10}$ MeV and ${\epsilon = 3\cdot 10^{-5}}$, 
{ X-particles can make all DM.} 
{Nevertheless  ${\Omega_X}$ wlll be taken as  free parameter.}

{The force from X-particles on electrons is 
$F = \sigma_{eX}\, v_{rel}\, n_X \,  m_X \,  v_{rot}$,
where
$ v_{rel}\sigma_{eX} = {4\pi \alpha \alpha' L}/{m_X^2 v_{rel}^3} $,
$m_X n_X = 10\, \Omega_X h^2 \kappa (z) (1+z)^3 \,{\rm keV}/{\rm cm}^3 $,
and {${\kappa (z)}$} is the dark matter overdensity in galactic halo with respect 
to the mean cosmological density at redshift ${z}$. 

Before discussing the generation of $B$ by $X$-particles let us comment on their role
in the structure formation. 
Prior to recombination the characteristic scattering time of light $X$-particles is shorter than the universe age, 
{${\tau_{Xe} < t_U}$}, so they are frozen in ${e\gamma}$-liquid.} After recombination and till reionization 
they behave as the usual WDM. 
{After reionization}  ${\tau_{Xe}}$  
again becomes shorter than $t_U$ 
and {the rotating ordinary matter in a protogalaxy would transfer a 
part of angular momentum to $X$-particles} and involve it in its turbulent motion.  So at this stage X-particles
behave similarly to the usual matter.

For an estimate of the magnitude of $B$ generated by light X-particles we use the obtained above equations but integrate
them till reionization, ${z=6}$ or  ${t_u = 1}$~Gyr. We take ${R= 100}$ kpc, ${\kappa = 100}$, ${v_{rot} = 10 }$ km/sec,
and impose the limit ${ \Omega_x h^2 = 0.007}$ to find $B =10^{-11}\,{\rm G}\, {\epsilon_5^2}/{m^2_{keV}}$.
{B can rise by factor 100, becoming ${ 10^{-9}}$ G,}
when the protogalaxy shrinks
from  100 kpc to 10 kpc, by far larger then the minimal necessary strength of the seed.

{For heavier X, ${m_X > m_e}$, a larger charge is allowed, ${\epsilon > 10^{-4}}$,} and
{X-particles can make all dark matter.}
After reionization, electron scatterings would not force X-particles into the galaxy rotation
and thus the effective integration time can be longer and  {magnetic fields as large as ${10^{-9}}$ G can be generated.}

To conclude, we have shown that an existence of millicharged particles with mass in keV - MeV range allows to:}\\
{1. Explain the origin of galactic and intergalactic magnetic fields.}\\
{2. Introduce DM with time dependent interaction with normal matter.} \\
{3. To solve or smooth down the problems of galactic satellites, 
angular momentum, and cusps in galactic centres  inherent to $\Lambda$CDM-cosmology.} \\
4. { The model can be tested in direct experiment. }

\section*{Acknowledgment}
The author acknowledge the support  of the Russian Federation Government Grant 
No. 11.G34.31.0047.


\end{document}